\documentclass[]{interact}

\usepackage{epstopdf}% To incorporate .eps illustrations using PDFLaTeX, etc.
\usepackage[caption=false]{subfig}% Support for small, `sub' figures and tables

\usepackage[natbibapa,nodoi]{apacite}
\usepackage[natbibapa,nodoi]{apacite}% Citation support using apacite.sty. Commands using natbib.sty MUST be deactivated first!
\theoremstyle{plain}% Theorem-like structures provided by amsthm.sty

\theoremstyle{definition}

\theoremstyle{remark}

\begin{document}

%\articletype{ARTICLE TEMPLATE}% Specify the article type or omit as appropriate

\title{Synthesis of memristive one-ports with piecewise-smooth characteristics}

\author{
\name{Vladimir~V. Semenov\textsuperscript{a,b}\thanks{CONTACT V.~V. Semenov. Email: semenov.v.v.ssu@gmail.com}}
\affil{\textsuperscript{a} Institut f\"{u}r Theoretische Physik, Technische Universit\"{a}t Berlin, Hardenbergstra{\ss}e 36, Berlin, Germany; 
\textsuperscript{b}Institute of Physics, Saratov State University, Astrakhanskaya str. 83, Saratov, Russia}
}

\maketitle

\begin{abstract}
A generalized approach for the implementation of memristive two-terminal circuits with piecewise-smooth characteristics is proposed on an example of a multifunctional circuit based on a transistor switch. Two versions of the circuit are taken into consideration: an experimental model of the piecewise-smooth memristor (Chua's memristor) and a piecewise-smooth memristive capacitor. Physical experiments are combined with numerical modelling of the discussed circuit models. Thus, it is demonstrated that the considered circuit is a flexible solution for synthesis of a wide range of memristive systems with tuneable characteristics. 
\end{abstract}


\section{Introduction}

The class of memristive systems unites a broad variety of dynamical systems of different nature. The significant feature of such systems is a continuous functional dependence of characteristics at any time on the previous states of the memristive system. Generally, memristive systems are described by the following equations (\cite{chua1976}):
\begin{equation}
 y=g(\boldsymbol{z},x,t)x,\quad
\dfrac{d\boldsymbol{z}}{dt}=\boldsymbol{f}(\boldsymbol{z},x,t),
\label{mem-sys}
\end{equation}
where $x$ is the input signal, $y$ is the system's response, $\boldsymbol{z}\in \mathbb {R}^{n}$ is a vector denoting the system state, $\boldsymbol{f}(\boldsymbol{z},x,t)$ is a continuous $n$-dimensional vector function, $g(\boldsymbol{z},x,t)$ is a continuous scalar function. It is assumed that the state equation  $\boldsymbol{\dot{z}}=\boldsymbol{f}(\boldsymbol{z},x,t)$ has a unique solution for any initial state $\boldsymbol{z_{0}}\in \mathbb {R}^{n}$. This equation reflects the dependence on the previous states as $z_{n}(t)=\int\limits_{-\infty}^{t}{f_{n}(\boldsymbol{z},x,t)dt}$. 

Memristive elements are promising candidates for a broad spectrum of applications including the development of next generation memory and computing systems (\cite{pershin2012,ventra2013,gaba2013,tetzlaff2014,vourkas2016,ielmini2016,wang2017}) and other issues involving both neurodynamics and electronics (\cite{kozma2012,adamatzky2014,jo2010,pershin2009,pershin2010-2}). The properties of memristive systems are frequently used for analog circuit development, such as programmable amplifiers, comparators and attenuators (\cite{pershin2010,shin2011}), adaptive filters (\cite{chew2012,ascoli2013}), to name only a few. Among a wide spectrum of meristor-based digital circuits one can distinguish encoders and converters which provide for implementing combinatorial circuits (\cite{singh2020,maruf2022}). The memristive systems attract attention of theorists as elements, whose intrinsic peculiarities can essentially change the dynamics of electronic oscillatory systems and are responsible for qualitatively new types of the behaviour: chaotic behaviour (\cite{buscarino2012,buscarino2013,pham2013,pham2015,chen2015}), the Hamiltonian dynamics (\cite{itoh2011}), the existence of hidden attractors (\cite{pham2015,chen2015}), the appearance of manifolds of equilibria and bifurcations without parameters (\cite{riaza2012,korneev2017,korneev2017-2,korneev2021}). 

Despite an exponentially increasing number of publications and growing interest of researchers, certain issues addressing the memristive systems are not studied in full. In particular, most of papers are focused on the memristor properties, while the features and applications of reactive memristive elements are explored in less details. Furthermore, the specificity of piecewise-smooth memristive characteristics is out of consideration in most cases. At the same time, there are many examples when the presence of piecewise-smooth nonlinearity is a significant feature being responsible for specific kinds of the oscillatory dynamics and bifurcation transitions (\cite{zhusubaliyev2003}). For this reason, the development of memristive one-port circuits with piecewise-smooth characteristic seems to be applicable in the context of practical applications and experimental studies of electronic oscillators including memristive circuits and components with piecewise-smooth nonlinearity.  

The proposed in the current letter electronic circuit is a particular solution where the state equation is electronically realised by means of analog modelling principles (\cite{luchinsky1998}). It provides for creation of a wide range of differential equations. Meanwhile, the circuit contains a transistor-based switch. The reactive and resistive properties of the circuit are fully determined by the transistor switch load. The combination of two mentioned approaches significantly extends a manifold of potentially implemented memristive devices.

\section{General description}
\begin{figure}[t]
\centering
\includegraphics[width=0.5\textwidth]{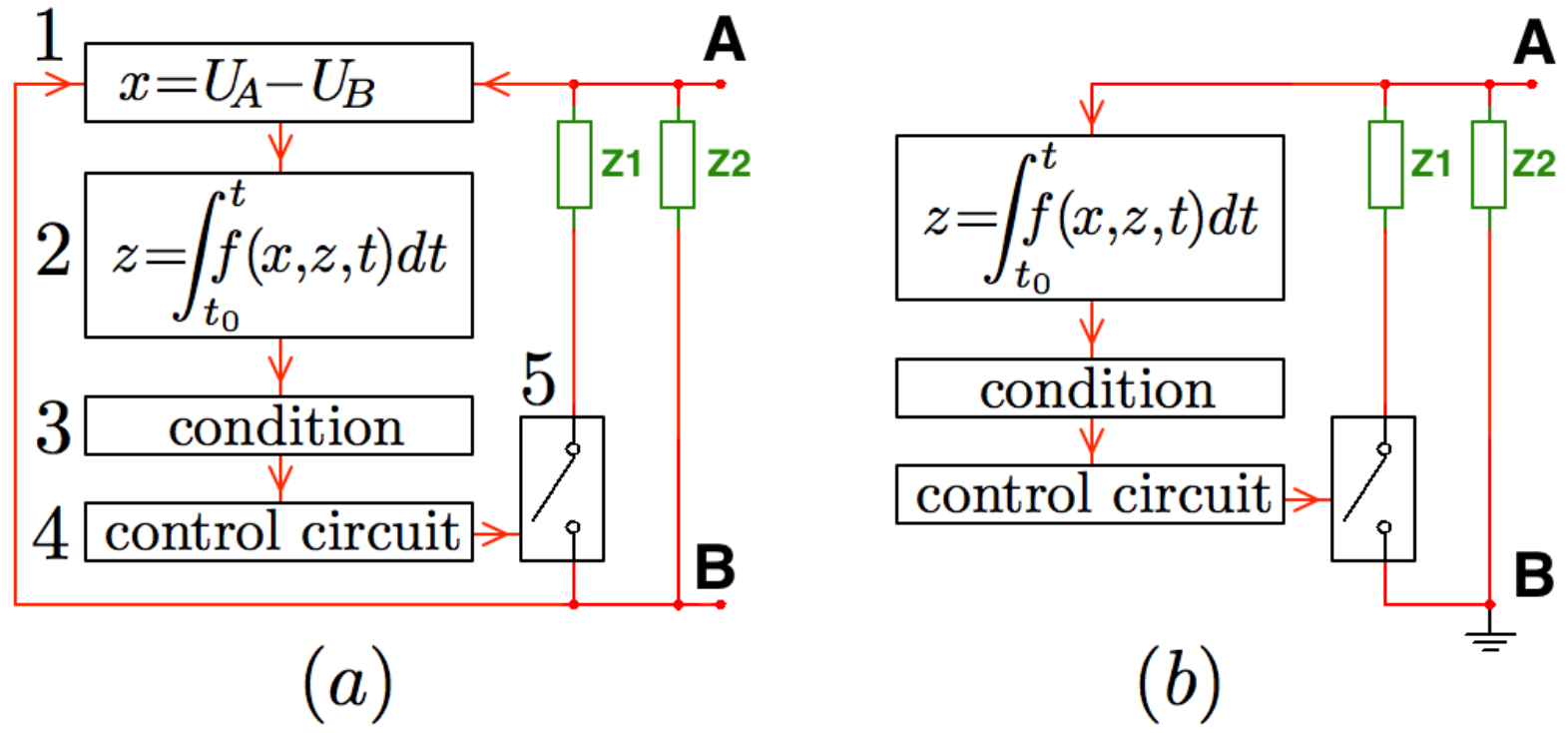}
\caption{Generalized scheme for memristive circuit synthesis. The difference between panels (a) and (b) consists in the presence of the reference point at terminal B.}
\label{fig1}
\end{figure}

The operation principle of the considered circuit is illustrated in Fig.\ref{fig1} (a). It includes two parallel one-ports Z1 and Z2. Element Z2 is always connected with terminals A and B. The connection of element Z1 with terminal B depends solely on switch circuit 5 controlled by feedback. The feedback consists of blocks 1-4. Block 1 produces a signal being equal to the voltage drop across the load (between terminals A and B): $x(t)=U_{A}(t)-U_{B}(t)$. If terminal B is connected with the ground point, block 1 can be excluded as in Fig. \ref{fig1} (b). Then $x(t)$ is the input signal for integrator 2, which forms the signal $z(t)$ according to the state equation of system (\ref{mem-sys}). After that, the signal $z(t)$ reaches the block of condition (block 3). If the condition is fulfilled, the corresponding signal comes to control circuit 4, which opens and closes switch 5. When switch 5 is closed, the circuit conductance is the sum of the conductances of elements Z1 and Z2. Otherwise, the switch is open and the conductance of element Z1 is excluded from the resulting circuit conductance. It is assumed that the controlling feedback has zero input current and has no time delay.
%%%%%%%%%%%%%%%%%%FIG1%%%%%%%%%%%%%%%%%%%%%%%%%%%%
%%%%%%%%%%%%%%%%%%%%%%%%%%%%%%%%%%%%%%%%%%%%%%%%%

%%%%%%%%%%%%%%%%%%%%%%%%%%%%%%%%%%%%%%%%%%%%%%%%%
\section{Memristor}
\begin{figure}[t]
\centering
\includegraphics[width=0.5\textwidth]{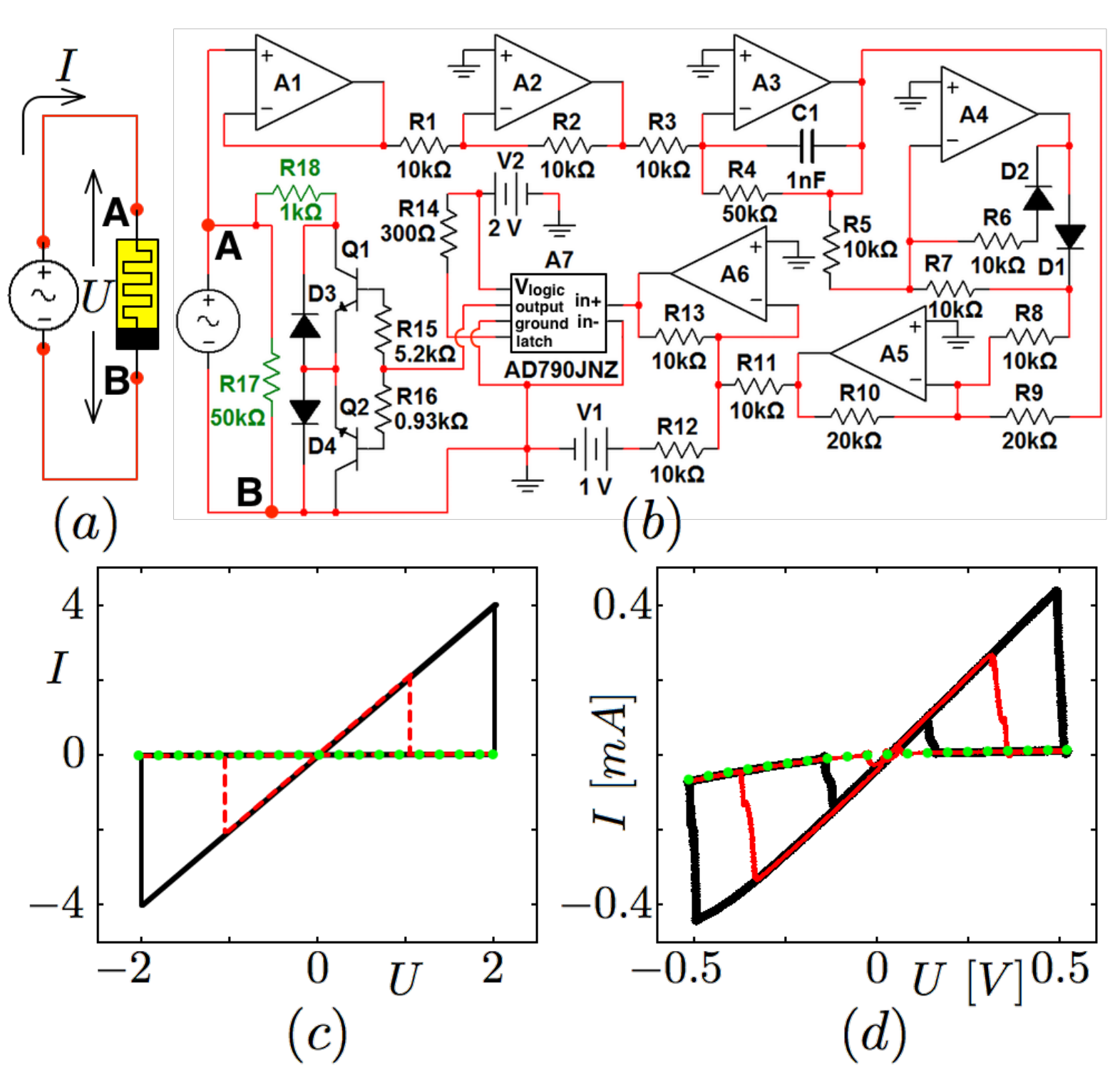}
\caption{Piecewise-smooth memristor under periodic forcing (panel (a)); (b) Circuit diagram of the particular memristor circuit implementation. Elements D1 and D2 are diodes 1N4148, D3 and D4 are diodes 1N4007, Q1 and Q2 are transistors 2N2222A, A1-A6 are operational amplifiers TL072CP; (c) Current-voltage characteristic of memristor model (\ref{memristor-num}) with $U=U_{max}\sin(2\pi f_{e}t)$. The parameters are: $a=0.02$, $b=2$, $\varphi_{0}=1$, $U_{max}=2$, $f_{e}=0.32$ (black solid curve), $f_{e}=0.59$ (red dashed curve), $f_{e}=0.65$ (green dotted curve); (d) Experimentally acquired characteristic of the experimental setup (\ref{memristor-exp}) (depicted in panel (b)) under external driving $U_{e}(t)=U_{max}\sin(2\pi f_{e}t)$. The parameters are: $U_{max}=0.5$ V, $f_{e}=4.0$ kHz (black solid line), $f_{e}=7.0$ kHz (red solid line), $f_{e}=8.1$ kHz (green dotted line).}
\label{fig2}
\end{figure}

The description of the memristive one-port implementation starts from the memristor electronic realization. One of the simplest memristor model was introduced by Leon Chua (\cite{chua1971}). In the dimensionless form its current-voltage characteristic is given by:
\begin{equation}
I=G_{M}(\varphi)U, \quad \dfrac{d\varphi}{dt}=U,\quad G_{M}(\varphi)=
\begin{cases}
          a , & |\varphi| \leq \varphi_{0},\\
          b , & |\varphi| > \varphi_{0},
\end{cases}
\label{memristor-num}
\end{equation}
where $I$ is the dimensionless current passing through the memristor, $U$ is the dimensionless voltage across the memristor, $\varphi$ is the dimensionless flux controlling the memristor state, the dimensionless parameter $G_{M}(\varphi)$ plays a role of the flux-controlled memristive conductance. In the presence of the external periodic signal $U_{e}(t)=U_{max}\sin{(2\pi f_{e} t)}$ [Fig.\ref{fig2} (a)], memristor model (\ref{memristor-num}) exhibits the current-voltage characteristic shaped as the pinched hysteresis loop [Fig.\ref{fig2} (c)] being typical for such systems (\cite{maruf2020}). When the frequency $f_{e}$ increases, the pinched hysteresis loop transforms into a single-valued function. 
%%%%%%%%%%%%%%%%%%%%%FIG2%%%%%%%%%%%%%%%%%%%%%%%%
%%%%%%%%%%%%%%%%%%%%%%%%%%%%%%%%%%%%%%%%%%%%%%%%%

The electronic circuit realization of system (\ref{memristor-num}) has been developed [Fig.\ref{fig2} (b)] according to the principles described above. The circuit has two terminals A and B. Element Z2 is a resistor ($R_{17}=50 k\Omega$) which is always connected with terminals A and B. The second load element, Z1, is a resistor ($R_{18}=1k\Omega$) connected between terminals A and B in parallel to Z2 when the electronic switch is closed. In this case the total resistance between terminals A and B is  $R_{\text{closed}}=\frac{R_{17}\times R_{18}}{R_{17}+R_{18}}\approx 980.4 \Omega$. When the electronic switch is open, the total resistance becomes $R_{\text{open}}=R_{17}=50k\Omega$. The electronic AC switch controlled by feedback consists of two transistors Q1 and Q2 connected with  diodes D3, D4  and resistors R15, R16.  The feedback consists of several blocks. The first element is analog follower A1 which output signal is denoted as $x=x(t)$ similarly to Fig. \ref{fig1} (a). Then the output signal of inverter A2 equals to $-x(t)$. The signal $-x$ comes to analog integrator A3, which is responsible for the differential equation defining the state variable $z$. The equation for integrator  A3 is $R_{3}C_{1}\frac{dz}{dt}=x-\frac{R_{3}}{R_{4}}z$, where $z(t)$ is the integrator output voltage. Then the condition denoting the piecewise-smooth characteristic is formed. Operational amplifiers A4 and A5 extract the absolute value of the $z(t)$ variable (the voltage at operational amplifier A5 output equals to $-|z(t)|$). Operational amplifier A6 has the summing amplifier configuration with the output voltage being equal to $|z(t)|-1$. This signal comes to comparator A7. If the signal $|z(t)|-1$ is greater than zero (it corresponds to condition $|z(t)|>1$), the corresponding signal is generated at the output of comparator A7, and the transistor-based switch becomes closed. The equations describing the circuit are:

\begin{equation}
\label{memristor-exp}
\begin{array}{l}
I=G_{M}(z)U, \quad R_{3}C_{1}\dfrac{dz}{dt}=U-\dfrac{R_{3}}{R_{4}}z, \\ 
G_{M}(z)=
\begin{cases}
          \dfrac{1}{R_{17}} , & |z| \leq 1 \text{ V}, \\
          \\
          \dfrac{1}{R_{17}}+\dfrac{1}{R_{18}} , & |z| > 1 \text{ V}.
\end{cases}
\end{array}
\end{equation}

It is important to explain a role of the term $-\dfrac{R_{3}}{R_{4}}z$ in the state equation of system (\ref{memristor-exp}). Let us assume that the term $-\dfrac{R_{3}}{R_{4}}z$ is absent. In such a case, the expression for the state variable takes the form $z(t)=\dfrac{1}{R_{3}C_{1}}\int_{0}^{t}U(t)dt$. If the circuit is driven by a periodic influence $U_{e}(t)$, then the voltage across the load is $U(t)=U_{e}(t)=U_{max}\sin{(2\pi f_{e} t)}$. The variable $z$ is expressed as $z(t)=-\dfrac{1}{R_{3}C_{1}2\pi f_{e}}\cos(2\pi f_{e}t)+K$ in the absence of the term $-\dfrac{R_{3}}{R_{4}}z$ in the state equation. It represents periodic oscillations with fixed mean value $K$. However, the behaviour observed in electronic experiments differs from the expected dynamics. Since the initial power-up moment, the variable $z$ increases (or decreases) and reaches its maximal (minimal) value limited by the power-supply voltage. This is due to the following facts. The equation for OA-based integrator does not take into consideration certain peculiarities of real operational amplifiers, such as a small but nonzero input current and capacitor's leakage currents. Moreover, the external periodic signal can slightly fluctuate or have a nonzero mean value. In order to stabilize the oscillations of the variable $z$ and make the dynamics less sensitive to random perturbations, resistor R4 being responsible for the term $-\dfrac{R_{3}}{R_{4}}z$ was introduced into the circuit.

Experimentally registered current-voltage characteristic [Fig.\ref{fig2} (d)] demonstrates all the phenomena observed in numerical model [Fig.\ref{fig2} (c)]. The quantitative difference consists in the frequency range, where the memristive peculiarities are exhibited. In practice, this range is defined by the ratings of $R_{3}$ and $C_{1}$.

%The ideal capacitor is described by the current-voltage relationship $I=C\frac{dU}{dt}$. 
\subsection{Memristive capacitor}
Similarly to the memristor, one can introduce a model of a piecewise-smooth capacitor, whose capacity is a piecewise-smooth function, $C_{M}(\varphi)$, similarly to the dependence $G_{M}(\varphi)$ of system (\ref{memristor-num}):
\begin{equation}
\begin{array}{l}
I=C_{M}(\varphi)\dfrac{dU}{dt}, \quad \dfrac{d\varphi}{dt}=U-k\varphi,\\
 C_{M}(\varphi)=
\begin{cases}
          a , & |\varphi| \leq \varphi_{0},\\
          b , & |\varphi| > \varphi_{0}.
\end{cases}
\end{array}
\label{memcapacitor-num}
\end{equation}
%%%%%%%%%%%%%%%%%%%%%FIG3%%%%%%%%%%%%%%%%%%%%%%%%
\begin{figure}[t]
\centering
\includegraphics[width=0.5\textwidth]{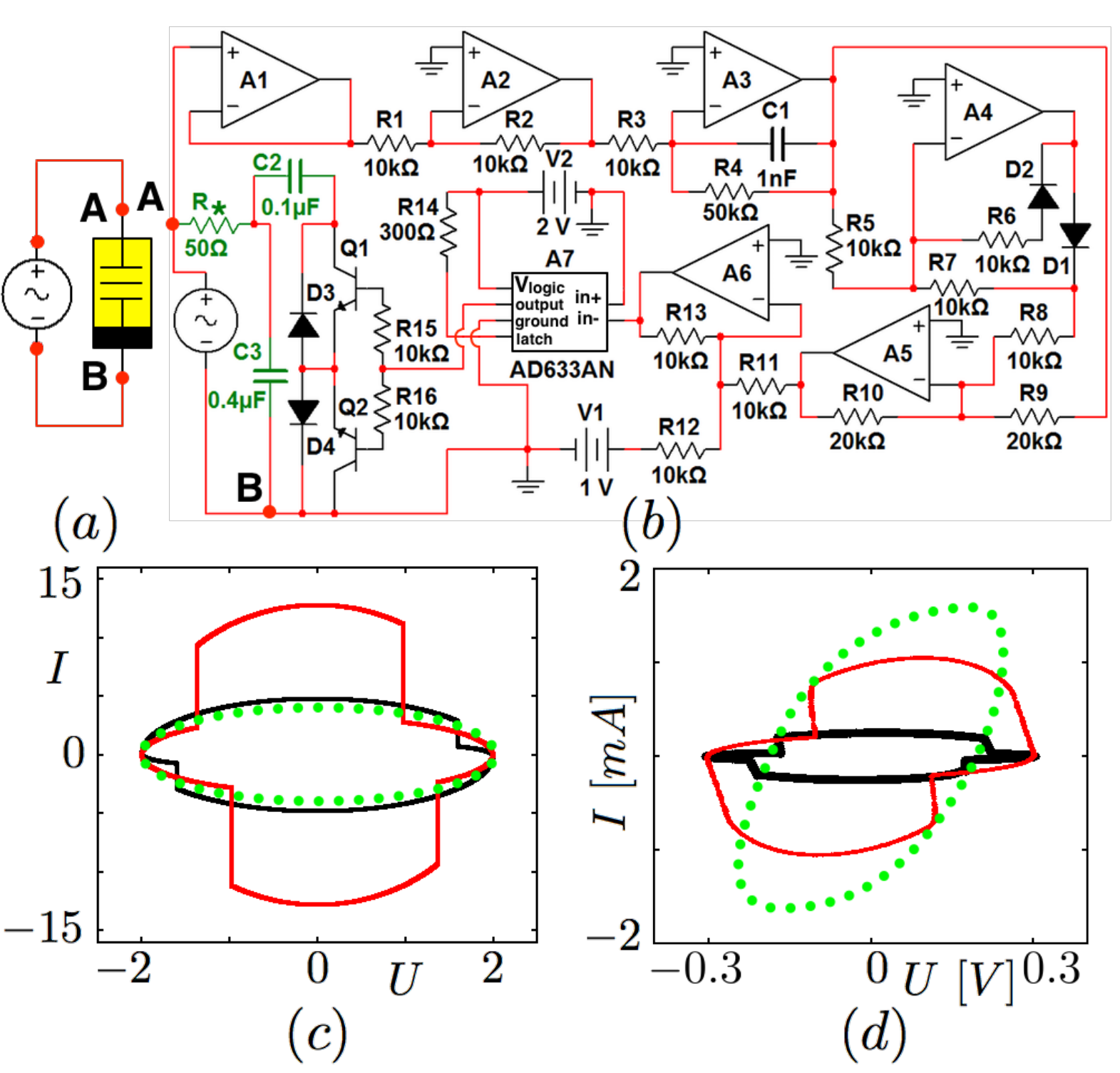}
\caption{Piecewise-smooth memcapacitor under periodic forcing (panel (a)); (b) Circuit diagram of the one-port circuit implementing memcapacitor. Elements D1, D2, D3, D4, Q, Q2, and A1-A7 are the same as in the previous figure; (c) Current-voltage characteristic of model (\ref{memcapacitor-num}) with $U=U_{max}\sin(2\pi f_{e}t)$. The parameters are: $k=0.2$, $a=1$, $b=4$, $\varphi_{0}=1$, $U_{max}=2$, $f_{e}=0.1$ (black solid curve), $f_{e}=0.25$ (red solid curve), $f_{e}=0.32$ (green dotted curve). (d) Experimentally registered characteristic of setup (\ref{memcapacitor-exp}) (depicted on panel (b)) under external driving $U_{e}(t)=U_{max}\sin(2\pi f_{e}t)$. The parameters are: $U_{max}=0.3$ V, $f_{e}=0.25$ kHz (black solid line), $f_{e}=1.3$ kHz (red solid line), $f_{e}=2.6$ kHz (green dotted line).}
\label{fig3}
\end{figure}
%%%%%%%%%%%%%%%%%%%%%%%%%%%%%%%%%%%%%%%%%%%%%%%%%
The memristive capacitor is assumed to be forced by the periodic signal $U=U_{max}\sin(2\pi f_{e}t)$ [Fig.\ref{fig3} (a)]. Then its current-voltage curve represents a closed circled curve which consists of parts corresponding to $C_{M}(\varphi)=a$ and $C_{M}(\varphi)=b$ [Fig.\ref{fig3} (c)]. Increasing $f_{e}$ leads to the transformation of the current-voltage curve into the ideal linear capacitor characteristic, $I=C\frac{dU}{dt}$. The experimental circuit being a model of the memcapacitor is depicted in Fig. \ref{fig3} (b).  The circuits illustrated in Fig. \ref{fig2} (b) and Fig. \ref{fig3} (b) have the same operating principle, but the transistor switch loads are different. The second difference between two circuits is the presence of the resistor $R_{*}=50\Omega$ in the memcapacitor circuit, which is necessary for correct functioning of the transistor switch. The following equations describe the circuit in Fig. \ref{fig3} (b) (the impact of the resistor $R_{*}$ is neglected):
\begin{equation}
\label{memcapacitor-exp}
\begin{array}{l}
I=C_{M}(z)\dfrac{dU}{dt}, \quad R_{3}C_{1}\dfrac{dz}{dt}=U-\dfrac{R_{3}}{R_{4}}z, \\ 
C_{M}(z)=
\begin{cases}
          C_{3} , & |z| \leq 1 \text{ V}, \\
          C_{2}+C_{3} , & |z| > 1 \text{ V}.
\end{cases}
\end{array}
\end{equation}

The current-voltage characteristic of model (\ref{memcapacitor-num}) and experimentally registered dependences $I(U)$ exhibit similar behaviour when the external driving frequency increases (compare Fig. \ref{fig3} (c) and Fig. \ref{fig3} (d)). The difference between the numerical model and the electronic device consists in the tilt of experimentally registered curve associated with the presence of the resistor $R_*$.

One can suppose that a memristive inductor can be successfully implemented in the similar way as the memcapacitor by introducing inductors as the switch load. However, it has been established that the development of the memristive inductor circuit is accompanied by additional difficulties associated with sharp impulses of the current through the load at the moments of switch opening and closing. Then the circuit operates incorrectly. Nevertheless, one can apply the circuits in Fig. \ref{fig2} (b) and Fig. \ref{fig3} (b) as memristive components of a gyrator and realize the memristive inductor properties. 

\section{Conclusion}
An electronic-circuit approach for the development of memristive one-ports has been demonstrated. The principal difference between the discussed method and a manifold of already published solutions (for instance, see the circuit implementations for the cubic memristor (\cite{muthuswamy2010}) and memcapacitor (\cite{yu2013}), the diode-bridge-based solution (\cite{bao2014}) and the analog emulator of the memristor with $\tanh$-nonlinearity (\cite{bao2019})) consists in the prospects for the piecewise-smooth nonlinearity realization. 

The advantage of the proposed circuit is its universality. Indeed, adjusting the circuit component ratings, one can control a frequency range, where the memristive properties are exhibited, as well as the resulting circuit impedance in different regimes. The proposed circuit solution allows for the implementation of any deterministic or stochastic forms of the memristor state equation. For instance, one can take into account the memristor forgetting effect (\cite{chang2011,chen2013,zhou2019}) or diffusive processes in the memristor (\cite{akther2021}) and experimentally study the manifestations of such phenomena. Moreover, when the switch is open or closed, the reactive properties of the load can be principally different (for example, elements Z1 and  Z2 can be a capacitor and a resistor, etc.). Thus, the described circuit can be applied for a broad variety of applications, including synthesis of memristive oscillators, adaptive filters, amplifiers and attenuators, etc. On the other hand, the considered circuits provide for experimental implementations of border-collision bifurcations (so-called C-bifurcations, see (\cite{zhusubaliyev2003})) observed in dynamical systems with piecewise-smooth nonlinearity. However, the discussed circuit demonstrates its own peculiarities. In particular, the experimental current-voltage characteristics become distorted for large amplitudes and low frequencies of the external forcing. This is due to the fact that the transistor-based switch does not work correctly for high current through the load.

\section*{Acknowledgements}
The author is very grateful to Anna Zakharova and Tatiana Vadivasova for helpful discussions and acknowledges support by the Russian Science Foundation (project No.  22-72-00038).

%\section{References}

%\begin{verbatim}
%%\bibliographystyle{apacite}
%%\bibliography{bibliography}
%\end{verbatim}

\end{document}